\documentclass[reprint, superscriptaddress, prl]{revtex4-2}

\usepackage{amsmath,amssymb}
\usepackage{graphicx}
\usepackage{bm}
\usepackage{times}
\usepackage{xcolor}
\usepackage{lineno}
\usepackage{textcomp, gensymb}
\usepackage{hyperref}

\newcommand{\uFive}{$^{235}$U}
\newcommand{\uEight}{$^{238}$U}
\newcommand{\pNine}{$^{239}$Pu}
\newcommand{\pOne}{$^{241}$Pu}

\newcommand{\STEREO}{\textsc{Stereo} }
\newcommand{\PROSPECT}{\textsc{Prospect} }

\begin{document}

\title{Joint Measurement of the \uFive \,Antineutrino Spectrum by PROSPECT and STEREO}

\newcommand{\MPIK}{\affiliation{Max-Planck-Institut f\"ur Kernphysik, Saupfercheckweg 1, 69117 Heidelberg, Germany}}
\newcommand{\LAPP}{\affiliation{Univ.~Grenoble Alpes, Universit\'e Savoie Mont Blanc, CNRS/IN2P3, LAPP, 74000 Annecy, France}}
\newcommand{\LPSC}{\affiliation{Univ.~Grenoble Alpes, CNRS, Grenoble INP, LPSC-IN2P3, 38000 Grenoble, France}}
\newcommand{\CEA}{\affiliation{IRFU, CEA, Universit\'e Paris-Saclay, 91191 Gif-sur-Yvette, France}}
\newcommand{\ILL}{\affiliation{Institut Laue-Langevin, CS 20156, 38042 Grenoble Cedex 9, France}}


\newcommand{\BU}{\affiliation{Department of Physics, Boston University, Boston, MA, USA}}
\newcommand{\BNL}{\affiliation{Brookhaven National Laboratory, Upton, NY, USA}}
\newcommand{\Drexel}{\affiliation{Department of Physics, Drexel University, Philadelphia, PA, USA}}
\newcommand{\GIT}{\affiliation{George W.\,Woodruff School of Mechanical Engineering, Georgia Institute of Technology, Atlanta, GA USA}}
\newcommand{\UH}{\affiliation{Department of Physics \& Astronomy, University of Hawaii, Honolulu, HI, USA}}
\newcommand{\IIT}{\affiliation{Department of Physics, Illinois Institute of Technology, Chicago, IL, USA}}
\newcommand{\NCSDLLNL}{\affiliation{Lawrence Livermore National Laboratory, Livermore, CA, USA}}
\newcommand{\LeMoyne}{\affiliation{Department of Physics, Le Moyne College, Syracuse, NY, USA}}
\newcommand{\MIT}{\affiliation{Department of Physics, Massachusetts Institute of Technology, Cambridge, MA, USA}}
\newcommand{\NIST}{\affiliation{National Institute of Standards and Technology, Gaithersburg, MD, USA}}
\newcommand{\HFIRORNL}{\affiliation{High Flux Isotope Reactor, Oak Ridge National Laboratory, Oak Ridge, TN, USA}}
\newcommand{\PDORNL}{\affiliation{Physics Division, Oak Ridge National Laboratory, Oak Ridge, TN, USA}}
\newcommand{\Temple}{\affiliation{Department of Physics, Temple University, Philadelphia, PA, USA}}
\newcommand{\UT}{\affiliation{Department of Physics and Astronomy, University of Tennessee, Knoxville, TN, USA}}
\newcommand{\Waterloo}{\affiliation{Institute for Quantum Computing and Department of Physics and Astronomy, University of Waterloo, Waterloo, ON, Canada}}
\newcommand{\UWis}{\affiliation{Department of Physics, University of Wisconsin, Madison, Madison, WI, USA}}
\newcommand{\Yale}{\affiliation{Wright Laboratory, Department of Physics, Yale University, New Haven, CT, USA}}

\author{H. Almaz\'an\ensuremath{^{\sigma}}}\altaffiliation[Present address: ]{Donostia International Physics Center, Paseo Manuel Lardizabal, 4, 20018 Donostia-San Sebastian, Spain}\MPIK
\author{M.\,Andriamirado\ensuremath{^{\pi}}}
\affiliation{Department of Physics, Illinois Institute of Technology, Chicago, IL, USA}
\author{A.\,B.\,Balantekin\ensuremath{^{\pi}}}
\affiliation{Department of Physics, University of Wisconsin, Madison, WI, USA}
\author{H.\,R.\,Band\ensuremath{^{\pi}}}
\affiliation{Wright Laboratory, Department of Physics, Yale University, New Haven, CT, USA}
\author{C.\,D.\,Bass\ensuremath{^{\pi}}}
\affiliation{Department of Physics, Le Moyne College, Syracuse, NY, USA}
\author{D.\,E.\,Bergeron\ensuremath{^{\pi}}}
\affiliation{National Institute of Standards and Technology, Gaithersburg, MD, USA}
\author{L. Bernard\ensuremath{^{\sigma}}}\altaffiliation[Present address: ]{Ecole Polytechnique, CNRS/IN2P3, Laboratoire  Leprince-Ringuet, 91128 Palaiseau, France}\LPSC
\author{A. Blanchet\ensuremath{^{\sigma}}}\altaffiliation[Present address: ]{LPNHE, Sorbonne Universit\'e, Universit\'e de Paris, CNRS/IN2P3, 75005 Paris, France}\CEA
\author{A. Bonhomme\ensuremath{^{\sigma}}}\MPIK\CEA
\author{N.\,S.\,Bowden\ensuremath{^{\pi}}}
\affiliation{Nuclear and Chemical Sciences Division, Lawrence Livermore National Laboratory, Livermore, CA, USA}
\author{C.\,D.\,Bryan\ensuremath{^{\pi}}}
\affiliation{High Flux Isotope Reactor, Oak Ridge National Laboratory, Oak Ridge, TN, USA}
\author{C. Buck\ensuremath{^{\sigma}}}\MPIK
\author{T.\,Classen\ensuremath{^{\pi}}}
\affiliation{Nuclear and Chemical Sciences Division, Lawrence Livermore National Laboratory, Livermore, CA, USA}
\author{A.\,J.\,Conant\ensuremath{^{\pi}}}
\affiliation{High Flux Isotope Reactor, Oak Ridge National Laboratory, Oak Ridge, TN, USA}
\author{G.\,Deichert\ensuremath{^{\pi}}}
\affiliation{High Flux Isotope Reactor, Oak Ridge National Laboratory, Oak Ridge, TN, USA}
\author{P. del Amo Sanchez\ensuremath{^{\sigma}}}\LAPP
\author{A.\,Delgado\ensuremath{^{\pi}}}
\affiliation{Physics Division, Oak Ridge National Laboratory, Oak Ridge, TN, USA} \affiliation{Department of Physics and Astronomy, University of Tennessee, Knoxville, TN, USA}
\author{M.\,V.\,Diwan\ensuremath{^{\pi}}}
\affiliation{Brookhaven National Laboratory, Upton, NY, USA}
\author{M.\,J.\,Dolinski\ensuremath{^{\pi}}}\affiliation{Department of Physics, Drexel University, Philadelphia, PA, USA}
\author{I. El Atmani\ensuremath{^{\sigma}}}\altaffiliation[Present address: ]{Hassan II University, Faculty of Sciences, A\"in Chock, BP 5366 Maarif, Casablanca 20100, Morocco}\CEA
\author{A.\,Erickson\ensuremath{^{\pi}}}
\affiliation{George W.\,Woodruff School of Mechanical Engineering, Georgia Institute of Technology, Atlanta, GA USA}
\author{B.\,T.\,Foust\ensuremath{^{\pi}}}
\affiliation{Wright Laboratory, Department of Physics, Yale University, New Haven, CT, USA}
\author{J.\,K.\,Gaison\ensuremath{^{\pi}}}
\affiliation{Wright Laboratory, Department of Physics, Yale University, New Haven, CT, USA}
\author{A.\,Galindo-Uribarri\ensuremath{^{\pi}}}\affiliation{Physics Division, Oak Ridge National Laboratory, Oak Ridge, TN, USA} \affiliation{Department of Physics and Astronomy, University of Tennessee, Knoxville, TN, USA}
\author{C.\,E.\,Gilbert\ensuremath{^{\pi}}}\affiliation{Physics Division, Oak Ridge National Laboratory, Oak Ridge, TN, USA} \affiliation{Department of Physics and Astronomy, University of Tennessee, Knoxville, TN, USA}
\author{S.\,Hans\ensuremath{^{\pi}}}\affiliation{Brookhaven National Laboratory, Upton, NY, USA}
\author{A.\,B.\,Hansell\ensuremath{^{\pi}}}
\affiliation{Department of Physics, Temple University, Philadelphia, PA, USA}
\author{K.\,M.\,Heeger\ensuremath{^{\pi}}}
\affiliation{Wright Laboratory, Department of Physics, Yale University, New Haven, CT, USA}
\author{B.\,Heffron\ensuremath{^{\pi}}}\affiliation{Physics Division, Oak Ridge National Laboratory, Oak Ridge, TN, USA} \affiliation{Department of Physics and Astronomy, University of Tennessee, Knoxville, TN, USA}
\author{D.\,E.\,Jaffe\ensuremath{^{\pi}}}
\affiliation{Brookhaven National Laboratory, Upton, NY, USA}
\author{S.\,Jayakumar\ensuremath{^{\pi}}}\affiliation{Department of Physics, Drexel University, Philadelphia, PA, USA}
\author{X.\,Ji\ensuremath{^{\pi}}}
\affiliation{Brookhaven National Laboratory, Upton, NY, USA}
\author{D.\,C.\,Jones\ensuremath{^{\pi}}}
\affiliation{Department of Physics, Temple University, Philadelphia, PA, USA}
\author{J. Koblanski\ensuremath{^{\pi}}}\affiliation{Department of Physics \& Astronomy, University of Hawaii, Honolulu, HI, USA}
\author{O.\,Kyzylova\ensuremath{^{\pi}}}\affiliation{Department of Physics, Drexel University, Philadelphia, PA, USA}
\author{L. Labit\ensuremath{^{\sigma}}}\LAPP
\author{J. Lamblin\ensuremath{^{\sigma}}}\LPSC
\author{C.\,E.\,Lane\ensuremath{^{\pi}}}\affiliation{Department of Physics, Drexel University, Philadelphia, PA, USA}
\author{T.\,J.\,Langford\ensuremath{^{\pi}}}
\affiliation{Wright Laboratory, Department of Physics, Yale University, New Haven, CT, USA}
\author{J.\,LaRosa\ensuremath{^{\pi}}}
\affiliation{National Institute of Standards and Technology, Gaithersburg, MD, USA}
\author{A. Letourneau\ensuremath{^{\sigma}}}\CEA
\author{D. Lhuillier\ensuremath{^{\sigma}}}\CEA
\author{M. Licciardi\ensuremath{^{\sigma}}}\LPSC
\author{M. Lindner\ensuremath{^{\sigma}}}\MPIK
\author{B.\,R.\,Littlejohn\ensuremath{^{\pi}}}
\affiliation{Department of Physics, Illinois Institute of Technology, Chicago, IL, USA}
\author{X.\,Lu\ensuremath{^{\pi}}}\affiliation{Physics Division, Oak Ridge National Laboratory, Oak Ridge, TN, USA} \affiliation{Department of Physics and Astronomy, University of Tennessee, Knoxville, TN, USA}
\author{J.\,Maricic\ensuremath{^{\pi}}}\affiliation{Department of Physics \& Astronomy, University of Hawaii, Honolulu, HI, USA}
\author{T. Materna\ensuremath{^{\sigma}}}\CEA
\author{M.\,P.\,Mendenhall\ensuremath{^{\pi}}}\affiliation{Nuclear and Chemical Sciences Division, Lawrence Livermore National Laboratory, Livermore, CA, USA}
\author{A.\,M.\,Meyer\ensuremath{^{\pi}}}\affiliation{Department of Physics \& Astronomy, University of Hawaii, Honolulu, HI, USA}
\author{R.\,Milincic\ensuremath{^{\pi}}}\affiliation{Department of Physics \& Astronomy, University of Hawaii, Honolulu, HI, USA}
\author{P.\,E.\,Mueller\ensuremath{^{\pi}}}\affiliation{Physics Division, Oak Ridge National Laboratory, Oak Ridge, TN, USA} 
\author{H.\,P.\,Mumm\ensuremath{^{\pi}}}
\affiliation{National Institute of Standards and Technology, Gaithersburg, MD, USA}
\author{J.\,Napolitano\ensuremath{^{\pi}}}
\affiliation{Department of Physics, Temple University, Philadelphia, PA, USA}
\author{R.\,Neilson\ensuremath{^{\pi}}}\affiliation{Department of Physics, Drexel University, Philadelphia, PA, USA}
\author{J.\,A.\,Nikkel\ensuremath{^{\pi}}}
\affiliation{Wright Laboratory, Department of Physics, Yale University, New Haven, CT, USA}
\author{S.\,Nour\ensuremath{^{\pi}}}
\affiliation{National Institute of Standards and Technology, Gaithersburg, MD, USA}
\author{J.\,L.\,Palomino\ensuremath{^{\pi}}}
\affiliation{Department of Physics, Illinois Institute of Technology, Chicago, IL, USA}
\author{H. Pessard\ensuremath{^{\sigma}}}\LAPP
\author{D.\,A.\,Pushin\ensuremath{^{\pi}}}\affiliation{Institute for Quantum Computing and Department of Physics and Astronomy, University of Waterloo, Waterloo, ON, Canada}
\author{X.\,Qian\ensuremath{^{\pi}}}
\affiliation{Brookhaven National Laboratory, Upton, NY, USA}
\author{J.-S. R\'eal\ensuremath{^{\sigma}}}\LPSC
\author{J.-S. Ricol\ensuremath{^{\sigma}}}\LPSC
\author{C. Roca\ensuremath{^{\sigma}}}\MPIK
\author{R. Rogly\ensuremath{^{\sigma}}}\CEA
\author{R.\,Rosero\ensuremath{^{\pi}}}
\affiliation{Brookhaven National Laboratory, Upton, NY, USA}
\author{T. Salagnac\ensuremath{^{\sigma}}}\altaffiliation[Present address: ]{Institut de Physique des deux Infinis de Lyon, CNRS/IN2P3, Univ. Lyon, Universit\'e Lyon 1, 69622 Villeurbanne, France}\LPSC
\author{V. Savu\ensuremath{^{\sigma}}}\CEA
\author{S. Schoppmann\ensuremath{^{\sigma}}}\altaffiliation[Present address: ]{University of California, Department of Physics, Berkeley, CA 
94720-7300, USA and
Lawrence Berkeley National Laboratory, Berkeley, CA 94720-8153, USA}\MPIK 
\author{M.\,Searles\ensuremath{^{\pi}}}
\affiliation{High Flux Isotope Reactor, Oak Ridge National Laboratory, Oak Ridge, TN, USA}
\author{V. Sergeyeva\ensuremath{^{\sigma}}}\altaffiliation[Present address: ]{Institut de Physique Nucl\'eaire Orsay, CNRS/IN2P3, 15 rue Georges Clemenceau, 91406 Orsay, France}\LAPP
\author{T. Soldner\ensuremath{^{\sigma}}}\ILL
\author{A. Stutz\ensuremath{^{\sigma}}}\LPSC
\author{P.\,T.\,Surukuchi\ensuremath{^{\pi}}}
\affiliation{Wright Laboratory, Department of Physics, Yale University, New Haven, CT, USA}
\author{M.\,A.\,Tyra\ensuremath{^{\pi}}}
\affiliation{National Institute of Standards and Technology, Gaithersburg, MD, USA}
\author{R.\,L.\,Varner\ensuremath{^{\pi}}}\affiliation{Physics Division, Oak Ridge National Laboratory, Oak Ridge, TN, USA} 
\author{D.\,Venegas-Vargas\ensuremath{^{\pi}}}\affiliation{Physics Division, Oak Ridge National Laboratory, Oak Ridge, TN, USA} \affiliation{Department of Physics and Astronomy, University of Tennessee, Knoxville, TN, USA}
\author{M. Vialat\ensuremath{^{\sigma}}}\ILL
\author{P.\,B.\,Weatherly\ensuremath{^{\pi}}}\affiliation{Department of Physics, Drexel University, Philadelphia, PA, USA}
\author{C.\,White\ensuremath{^{\pi}}}
\affiliation{Department of Physics, Illinois Institute of Technology, Chicago, IL, USA}
\author{J.\,Wilhelmi\ensuremath{^{\pi}}}
\affiliation{Wright Laboratory, Department of Physics, Yale University, New Haven, CT, USA}
\author{A.\,Woolverton\ensuremath{^{\pi}}}\affiliation{Institute for Quantum Computing and Department of Physics and Astronomy, University of Waterloo, Waterloo, ON, Canada}
\author{M.\,Yeh\ensuremath{^{\pi}}}
\affiliation{Brookhaven National Laboratory, Upton, NY, USA}
\author{C.\,Zhang\ensuremath{^{\pi}}}
\affiliation{Brookhaven National Laboratory, Upton, NY, USA}
\author{X.\,Zhang\ensuremath{^{\pi}}}
\affiliation{Nuclear and Chemical Sciences Division, Lawrence Livermore National Laboratory, Livermore, CA, USA}

\collaboration{\ensuremath{^{\pi}}\PROSPECT and \ensuremath{^{\sigma}}\STEREO Collaborations}
\homepage{http://prospect.yale.edu, email: prospect.collaboration@gmail.com}
\homepage{http://www.stereo-experiment.org, email: stereo.collaboration@gmail.com}

\date{\today}
\begin{abstract}
    The \PROSPECT and \STEREO collaborations present a combined measurement of the pure \uFive{} antineutrino spectrum, without site specific corrections or detector-dependent effects.
    The spectral measurements of the two highest precision experiments at research reactors are found to be compatible with $\chi^2/\mathrm{ndf} = 24.1/21$, allowing a joint unfolding of the prompt energy measurements into antineutrino energy. 
    This $\bar{\nu}_e$ energy spectrum is provided to the community, and an excess of events relative to the Huber model is found in the 5-6~MeV region. 
    When a Gaussian bump is fitted to the excess, the data-model $\chi^2$ value is improved, corresponding to a $2.4\sigma$ significance.
    
\end{abstract}

\maketitle


Reactor-based experiments have played a fundamental role in developing our understanding of neutrinos~\cite{Cowan103,KamLAND,araki2005experimental}. 
The majority of these efforts have utilized nuclear power reactors due to their high overall antineutrino flux.
However, power reactors are not without drawbacks: experiments cannot easily be located close to the reactor and their low-enriched uranium (LEU) cores have a time-evolving mixture of fissioning isotopes.
Notably, the precision reactor-based $\theta_{13}$ experiments have all observed deviations between measured and predicted antineutrino fluxes and spectra \cite{PhysRevLett.123.111801,DC2020,Seo:2016uom} that point to deficiencies in the leading theoretical models \cite{RAA1,Huber} for the four primary fissioning isotopes (\uFive{}, \pNine{}, \pOne{}, \uEight{}).
The origin of these LEU-based discrepancies remains unknown, though there are indications that one or more fissioning isotopes have incorrectly predicted $\bar{\nu}_e$ fluxes and spectra \cite{PhysRevLett.118.251801,Bak_2019,Giunti_2017a,Giunti_2017b,Gebre_2018}.

A small number of experiments have targeted compact-core research reactors, where highly enriched uranium (HEU) cores are dominated by a single fissioning isotope, \uFive{}. 
These experiments have provided unique tools to search for short baseline neutrino oscillations and measure the \uFive{} antineutrino flux and spectrum \cite{ILL,SavannahRiver,Vidyakin:1987ue,Vidyakin:1990iz,Vidyakin:1994ut}.
Two recent HEU-based efforts are the \STEREO \cite{STEREODetector} and \PROSPECT \cite{Ashenfelter:2018zdm} experiments that have separately published searches for short-baseline neutrino oscillations and measurements of their detected antineutrino spectra \cite{STEREOOscillationPhase1and2,STEREORatePhase2,STEREOShapePhase2,Andriamirado_2021}.
Thanks to the large statistics available and accurate control of energy reconstruction, \PROSPECT and \STEREO provide an opportunity to study the \uFive{} antineutrino spectrum with unprecedented precision. 
Previous analyses by each collaboration indicate the presence of shape distortion for this specific isotope \cite{Ashenfelter:2018jrx,STEREOShapePhase2,Andriamirado_2021} without any assumption regarding spectra from the other fissioning isotopes.

In this Letter, these collaborations  present a combined analysis that leverages independent statistics and complementary detector technologies to reduce the effect of systematic uncertainties and produce a robust \uFive{}
antineutrino energy spectrum that can be used by current and future reactor experiments.
We also present comparisons of this spectrum to Huber's theoretical model for the \uFive{} spectrum \cite{Huber} and search for spectral deviations similar to those observed in the LEU experiments.


    




The \PROSPECT and \STEREO experiments are both operated near research reactors.
\PROSPECT is located within the HFIR facility at Oak Ridge National Laboratory, and \STEREO experiment near the R\'eacteur \`a  Haut Flux (RHF) of the ILL research centre in Grenoble, France. 
HFIR (RHF) uses an 85~MW$_\mathrm{th}$ (58.3~MW$_\mathrm{th}$) compact core with 93\% \uFive{} enriched fuel that operates with a 24-day (45-day) reactor-on cycle.
More than 99\% of antineutrino flux at the detector sites comes from \uFive{} fissions, with small non-fuel contributions coming from activated $^{28}$Al and $^6$He ($^{28}$Al and $^{55}$Mn) in structural materials.

Electron antineutrinos are detected through the inverse beta decay (IBD) process $\bar{\nu}_e + p \rightarrow \beta^+ + n$, which produces a pair of signals correlated in time and space. 
The prompt signal comes from the $\beta^+$ scintillation and annihilation in the detector medium and carries information about the $\bar{\nu}_e$ energy, while the delayed signal corresponds to the neutron capture after its thermalization. 
This pair structure is used to select signal events.

\textsc{Prospect}'s and \textsc{Stereo}'s detector sites are located in close proximity (about 10 meters) to the reactor cores in high background environments. 
For both experiments, reactor-induced background is mitigated by a passive shielding of lead, polyethylene, or borated polyethylene. 
The water channel of ILL's reactor building provides 15~m water equivalent shielding against cosmic rays for the \STEREO detector, while only about 0.5~m water equivalent is available for the \PROSPECT detector.

The \PROSPECT detector is a $\sim$4~ton active $^6$Li-loaded liquid scintillator (LiLS) target separated into 154 segments by reflector panels with enclosed photomultiplier tubes (PMTs) on each end~\cite{Ashenfelter:2018zdm}. 
The high segmentation of the detector compensates for the mild cosmic shielding by allowing background suppression from fiducialization, cosmic vetoing, and position reconstruction. 
In combination with the pulse shape discrimination (PSD) capability of the LiLS, it provides a signal-to-correlated background ratio of 1.4 in the signal energy range.

The target volume of the \STEREO detector is optically segmented in six identical cells filled with 1.6~tons liquid scintillator with neutron capture enhanced by Gd doping (GdLS). 
The scintillation light is read out by PMTs placed on top of the cells. 
An outer crown, segmented in four cells and filled with 1.85 tons unloaded liquid scintillator, mitigates energy leakage out of the target volume and improves background rejection \cite{STEREODetector}. 
With this, combined with an active muon-veto on top of the detector and PSD, a signal-to-correlated background ratio of 1.1 is achieved. 
Even if more shielding against cosmic rays were available, \textsc{Stereo}'s signal-to-background ratio is limited by its PSD capabilities and RHF's lower operational power.

Accurate knowledge of the detector energy response is a key aspect for both collaborations. 
A variety of calibration sources are used to ensure consistent measurement of the detected visible energy.
Extensive calibrations are regularly carried out with point-like $\gamma$ sources and neutron sources, circulated in and around the detectors. 
The control of both detector responses is further improved by the study of the $\beta$-decays of $^{12}$B atoms generated throughout the detectors by the interaction of cosmic rays.
These calibration efforts allow for full characterization of the \PROSPECT and \STEREO detectors and scintillators, which when combined with geometric modelling and simulation, allows the construction of a detector response model which reproduces observed effects in prompt energy to within 1\%.
 
This analysis uses the latest results of the \PROSPECT and \STEREO experiments as inputs.
The \PROSPECT (\textsc{Stereo}) spectrum measurement corresponds to 82 (119) live-days of exposure yielding 50560~$\pm$~406 (stat) (43400~$\pm$~382 (stat)) IBD candidates after background subtraction.

Uncertainties in the \PROSPECT and \STEREO spectrum measurements can be categorized as data effects, model effects, and detector effects.
Statistical uncertainties make up the majority of the data effects, but also included are systematic uncertainties in background rates.
Model uncertainties are related to the normalization of non-fuel corrections ($^{28}$Al and $^6$He for HFIR, $^{28}$Al and $^{55}$Mn for RHF) as well as non-equilibrium isotope corrections.
The remaining category (detector effects) include knowledge of the physical properties of the detectors that affect their response. 
For \textsc{Prospect}, such uncertainties are on nonlinearity and energy loss, as well as selection cuts for muon veto variations, fiducial volume, and energy thresholds. 
For \textsc{Stereo}, these are uncertainty on the energy scale, and uncertainty induced by selection cuts.
Covariance matrices are generated for each uncertainty, and combined to produce the full uncertainty of the measurement \cite{Andriamirado_2021,STEREOShapePhase2}.

The inputs of the analysis (prompt energy spectrum, response matrix, covariance matrix) for \PROSPECT can be found in \cite{Andriamirado_2021}. 
For the current work, two changes were made that differ from these inputs.
The first is shifting the neutrino energy binning of the \PROSPECT response matrix by 50~keV. 
This was necessary to match the binning convention of the joint analysis. 
The second change was reducing the uncertainty on the $^{28}$Al, $^6$He, and non-equilibrium contributions to the HFIR spectrum from 100\% to 25\%, going from a very conservative estimate to a more reasonably conservative estimate based on the original study of the contributions.
All relevant inputs for \STEREO can be found in the \texttt{HEPData} repository \cite{STEREOdatashare-shape} related to the recent publication \cite{STEREOShapePhase2} \nocite{STEREODOI1} \nocite{STEREODOI2} \nocite{STEREODOI3} \nocite{STEREODOI4}.






The prompt energy spectra published by both collaborations \cite{Andriamirado_2021,STEREOShapePhase2} use different energy responses and cannot be compared directly.
For instance, \STEREO has a prompt energy scale that includes a quenching of about 10\%, which is not corrected but rather reproduced with \%-level accuracy in the simulation (hence included in the detector response matrix). 
\PROSPECT reproduces an absolute energy scale in simulation from multiple calibration campaigns, but also has notable factors such as energy leakage and missing segments which affect the energy response.
Therefore, these spectral measurements must be mapped to a common energy space to be properly compared. 

\begin{figure}[tb]
   \includegraphics[width=\linewidth]{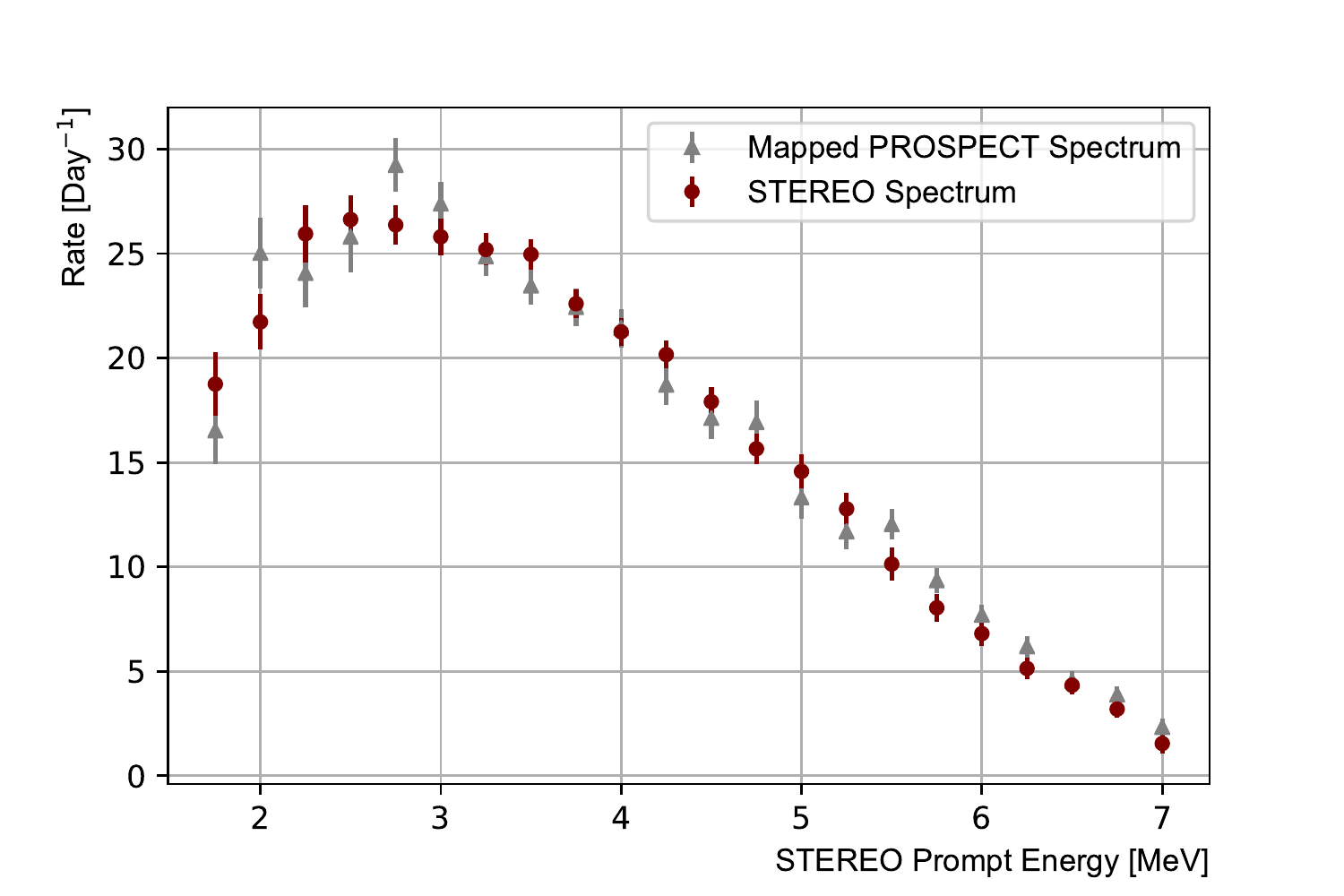}
    \caption{Comparison of \PROSPECT and \STEREO spectral measurements in the \STEREO prompt energy space showing good agreement.
    The \PROSPECT measurement has been mapped into the \STEREO prompt space and fit for normalization.
    }
    \label{fig1}
\end{figure}

Such a comparison is done by mapping one measurement into the prompt energy space of the other, i.e. unfolding the first one using the pseudo-inverse of its response matrix, then folding through the response matrix of the other. 
As the \PROSPECT measurement covers a wider analysis range in antineutrino energy, the \PROSPECT prompt energy spectrum was mapped into \STEREO prompt energy space. 
The two spectra are displayed in Figure~\ref{fig1}.
The comparison of both spectra in this prompt energy space, including a free-floating normalization parameter, gives $\chi^2/\mathrm{ndf} = 24.1/21$ (p-value: 0.29) and indicates that the two measurements are compatible and performing a joint measurement is relevant. 

In the following, the superscripts PR and ST respectively label \PROSPECT and \STEREO data. We assume the \PROSPECT and \STEREO measurement uncertainties are independent.
In this analysis the two prompt spectra, referred to as $D^\mathrm{PR}$ and $D^\mathrm{ST}$, are simultaneously unfolded into a single \uFive{} antineutrino energy spectrum denoted $\Phi^\mathrm{U5}$, with the respective non-\uFive{} flux corrections removed by subtracting the amount present in the prompt prediction. 
The resulting spectrum is reported in 250-keV wide antineutrino energy bins with bin centers ranging from 2.5 to 7.75~MeV.

Spectral unfolding is non-trivial, as inverting detector response matrices associated with each experiment, $R^\mathrm{PR}$ and $R^\mathrm{ST}$, induces some loss of information through smearing and statistical fluctuations are amplified.
To mitigate the increase in the variance of the unfolded spectrum, regularized unfolding techniques are used.
For more robustness in the final result, two complementary regularization approaches are presented.

The first unfolding technique is the Tikhonov regularization~\cite{tikhonov} where the $\chi^2$ to be minimized is given by:
\begin{eqnarray}\label{eqn:chi2-covmat}
\chi^2 \left(\beta, \Phi^\mathrm{U5}\right) &=& \boldsymbol{\Delta}\left(\beta, \Phi^\mathrm{U5}\right)^T \, V^{-1} \,  \boldsymbol{\Delta}\left(\beta, \Phi^\mathrm{U5}\right) \nonumber \\
&+& \mathcal{R}_{1}\left(\Phi^\mathrm{U5}\right)
\end{eqnarray}
where $\boldsymbol{\Delta}(\beta, \Phi^\mathrm{U5}) =  D^\mathrm{JNT} - R_\beta \Phi^\mathrm{U5}$. Here $D^\mathrm{JNT}$, $V$ and $R_\beta$ are respectively the joint prompt data (with non-$^{235}$U corrections subtracted beforehand), experimental covariance matrix accounting for statistical and systematic uncertainties, and detector response matrix:
\begin{eqnarray}
    D^\mathrm{JNT} = 
    \begin{pmatrix}
        D^\mathrm{PR}\\
        D^\mathrm{ST}
    \end{pmatrix}&,&\;
       V = 
    \begin{pmatrix}
        V^\mathrm{PR} & 0\\
       0 & V^\mathrm{ST}
    \end{pmatrix}
    ,\nonumber \\
        R_\beta &= &
    \begin{pmatrix}
       \beta \cdot R^\mathrm{PR}\\
        R^\mathrm{ST}
    \end{pmatrix},
    \label{eqn 2}
\end{eqnarray}
where $\beta$ is a free-floating scale parameter accounting for the difference in normalization of \PROSPECT and \STEREO data. 
The penalty term \begin{equation}\label{eqn:Tikh-reg}
    \mathcal{R}_{1}\left(\Phi^\mathrm{U5}\right) = r \sum_{i} \left( \frac{\Phi^\mathrm{U5}_{i+1}}{\Phi^\mathrm{H}_{i+1}} - \frac{\Phi^\mathrm{U5}_{i}}{\Phi^\mathrm{H}_{i}} \right)^2 
\end{equation}
with $r>0$ is a regularization term constraining the first derivative of the shape of the spectrum with respect to a prior shape $\Phi^\mathrm{H}$, set to be the \uFive{} Huber model from Ref.~\cite{Huber}, area-normalized to the data. 
Using the Generalized Cross-Validation prescription (GCV) to minimize the prediction error of the fit~\cite{GCV}, a regularization strength of $r=49$ is determined. 

The unfolded spectrum can then be expressed as $\Phi^\mathrm{U5} = H_{r,\beta} \cdot D^\mathrm{JNT}$ where
\begin{equation}\label{eqn:Tikhonov-matrix}
H_{r,\beta} = \left[ 1+r( R_\beta^T V^{-1} R_\beta)^{-1} C_T\right]^{-1} \left( R_\beta^T V^{-1} R_\beta\right)^{-1} R_\beta^T V^{-1}
\end{equation} is the regularized unfolding matrix, $C_T$ comes from rendering Eqn.~\ref{eqn:Tikh-reg} in matrix notation: $ \mathcal{R}_{1}(\Phi^\mathrm{U5})\equiv r \cdot (\Phi^\mathrm{U5})^T \cdot C_T \cdot (\Phi^\mathrm{U5})$, and $\beta$ is set to minimize the $\chi^2$ of Eqn.~\ref{eqn:chi2-covmat}.The covariance matrix of the spectrum $\Phi^\mathrm{U5}$ is then evaluated as $V_{\Phi} = H_{r,\beta}\, V \,H_{r,\beta}^T$.

The second technique used in this analysis is Wiener-SVD unfolding, which is optimized for the expected signal-to-background across the analysis range \cite{Tang_2017}.
Instead of using a regularization variable which must be tuned, the strength of the effective regularization is handled by the Wiener filter, $W_C$.
The unfolded data is prescribed as:
\begin{equation}\label{eqn:wiener-svd}
    \Phi^\mathrm{U5} = C^{-1} \cdot V_C \cdot W_C \cdot V_C^T \cdot C \cdot ( \widetilde{R}^T \widetilde{R})^{-1} \cdot \widetilde{R}^T \cdot \widetilde{D}^\mathrm{JNT}
\end{equation}
where $C$ is the 2nd derivative curvature matrix, $V_C$ is the right matrix from a singular value decomposition of $\widetilde{R} \cdot C$ and $W_C$ is the Wiener filter, which is used to extract the true signal while suppressing noise.
Here the response $\widetilde{R}$ is the stacked, pre-scaled responses of the separate experiments:
\begin{equation}
    \widetilde{R} = 
    \begin{pmatrix}
        \widetilde{R}^\mathrm{PR}\\
        \alpha \cdot \widetilde{R}^\mathrm{ST}
    \end{pmatrix}
    ,\;
\end{equation}
where $\alpha$ matches the relative scaling in \ref{eqn 2}.
Pre-scaling is done by applying the lower triangular matrix from the Cholesky decomposition of the covariance matrix.
The unfolded covariance matrix is generated by sampling unfolded prompt fluctuated toy spectra and comparing to the model.
Because the Wiener-SVD method requires estimation of the true model, an additional 3\% bin-to-bin model uncertainty is added to the toys since the exact fine structure of the spectrum is unknown, similar to Ref. \cite{DayaBay:2021dqj}.


We assessed the ability of each framework to retrieve a reference antineutrino energy spectrum shape. 
We generate joint pseudo-data by folding a reference spectrum into \PROSPECT and \STEREO prompt spaces, with the respective flux normalizations applied, and fluctuate these prompt spectra within both experimental uncertainties. 
Then, the joint unfolding of $10^4$ pairs of pseudo-data are used to compute biases induced by the unfolding methods. 
This bias study was performed for each framework, alternatively setting as the reference spectrum the Huber model $\Phi^\mathrm{H}$ or distorted models featuring a Gaussian event excess 
comparable to measurements reported by the Daya Bay \cite{PhysRevLett.123.111801} and \STEREO \cite{STEREOShapePhase2} collaborations. 
The relative bias is no more than 0.5\% for the Tikhonov framework and 1\% for the Wiener-SVD framework in the whole analysis range.

\begin{figure}[tb]
   \includegraphics[width=\linewidth]{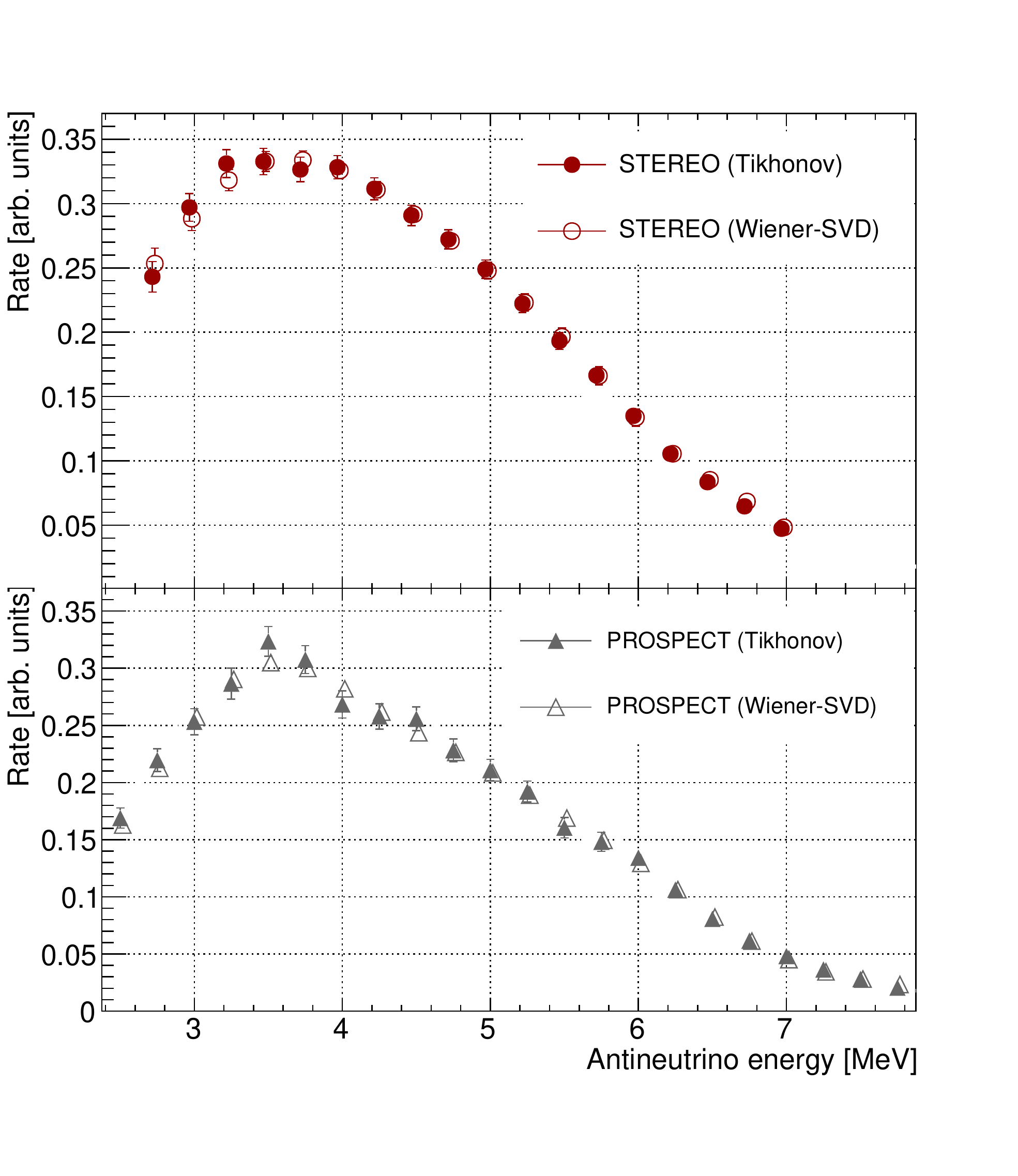}
    \caption{Unfolded $^{235}$U spectra from \STEREO (top panel) and \PROSPECT (bottom panel) data, using both frameworks.
    Error bars are the diagonal elements of the covariance matrix only. All spectra are normalized to unit area.}
    \label{fig2}
\end{figure}

The agreement between the frameworks can be illustrated with the separate unfolding of \PROSPECT and \STEREO measurements. 
The unfolded spectra are displayed in Figure~\ref{fig2}. 
Note that the unfolding of \STEREO data uses a restricted range due to requiring a selection efficiency $> 50\%$ \cite{STEREOShapePhase2}. 
One main difference between the two procedures consists in the level of smoothness that is applied on the spectrum during the unfolding. 
It can be seen comparing the unfolding of \PROSPECT data by both methods (full and empty triangles) in Figure~\ref{fig2}. 
The reader can observe that fluctuations in the 3.5-4.5~MeV region are preserved by the Tikhonov method, while being smoothed out by the Wiener-SVD method.

In general, each unfolding process will have its own level of induced  bias, smoothing and bin-to-bin correlations 
when transforming the prompt measurements to a spectrum in antineutrino energy.
These effects complicate interpretation of the unfolded spectrum but are necessary to control the amplification of statistical fluctuations in the measurement.
To account for this, a filter matrix, $A_C$, should be applied to models to make an accurate comparison with unfolded data \cite{Tang_2017}.
This matrix connects the regularized unfolding $\Phi_\mathrm{reg}$ (reported in this Letter) with the unregularized unfolding $\Phi_\mathrm{unreg}$ as
\begin{equation}
    \Phi_\mathrm{reg} = A_C \cdot \Phi_\mathrm{unreg}.
\end{equation}
The exact form of the $A_C$ matrix in the Wiener-SVD approach is $C^{-1} \cdot V_C \cdot W_C \cdot V_C^T \cdot C$ as described in Ref. \cite{Tang_2017}, whereas it reads $[ 1+r( R_\beta^T V^{-1} R_\beta)^{-1} C_T]^{-1}$ in the Tikhonov formalism. Note that these expressions appear as prefactors in the respective unfolding formulas (\ref{eqn:wiener-svd}) and (\ref{eqn:Tikhonov-matrix}) and thus factor out the regularization-dependent part of the unfolding.
Both filter matrices are included in the supplementary materials.

The two frameworks produce compatible jointly unfolded antineutrino spectrum results and well-controlled bias in comparison to models. 
Due to introducing less bias and using a more straightforward regularization method, the unfolding framework using Tikhonov regularization with strength tuned by the GCV prescription is used to present the main results of the unfolded joint spectrum and model comparisons in this Letter.
However, since the Wiener-SVD technique offers additional benefits outside the scope of this paper, such as optimized signal-to-background for the analysis energy range, the results of this framework are also included in the supplemental materials.
For general purposes, we recommend readers use the Tikhonov jointly unfolded result.

\begin{figure}[tb]
   \includegraphics[width=\linewidth]{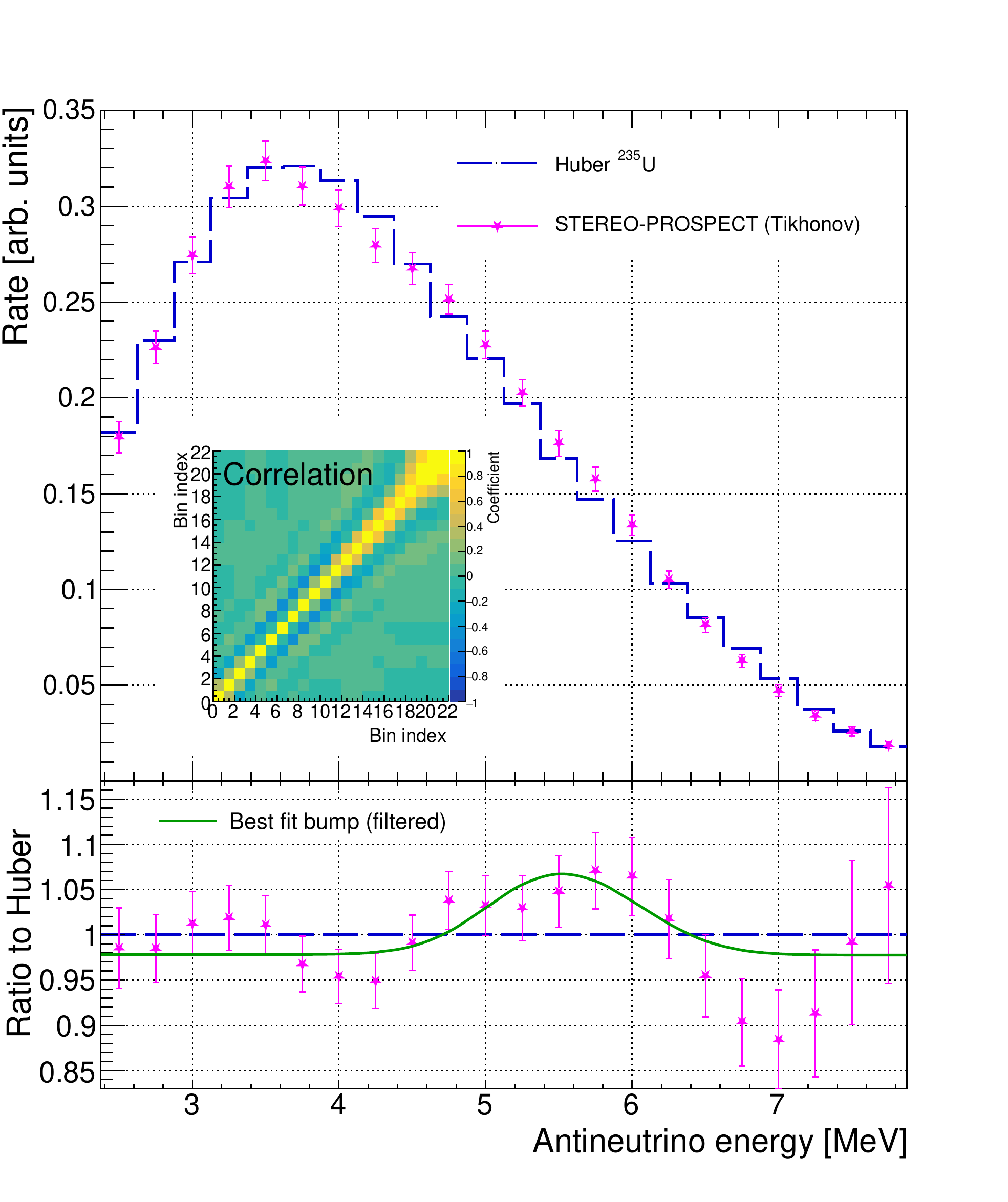}
    \caption{(Top) Jointly unfolded $^{235}$U spectrum with diagonal errors and Huber prediction normalized to unit area.
    The non-trivial correlation matrix is displayed. 
    (Bottom) Jointly unfolded $^{235}$U spectrum, as a ratio to Huber. 
    The filtered best-fit bump is displayed. 
    }
    \label{fig3}
\end{figure}

The comparison of the jointly unfolded $^{235}$U spectrum to the area-normalized Huber model \cite{Huber} is shown in Figure~\ref{fig3}. 
The $\chi^2$ comparison gives $\chi^2/\mathrm{ndf} = 30.8/21$. 
A localized event excess is found in the 5-6 MeV region in antineutrino energy. 
This excess with respect to the Huber model $\Phi^\mathrm{H}$ can be described by a Gaussian, and the following model with 4 free parameters:
\begin{equation}
    M(E_\nu) = a \cdot \Phi^\mathrm{H}(E_\nu) \left[ 1+ A \exp -\frac{(E_\nu - \mu)^2}{2\sigma^2} \right]
\end{equation} is fitted against the joint spectrum $\Phi^\mathrm{U5}$. 
The global normalization parameter $a$ ensures a shape-only comparison.
To account for unfolding biases, the fit is performed through the filter matrix as 
\begin{equation}\label{eqn:gaus-fit}
    \chi^2 = \big(A_C \cdot M - \Phi^\mathrm{U5})^T V^{-1}_\Phi \big(A_C \cdot M - \Phi^\mathrm{U5}).
\end{equation}
The result is displayed in the bottom panel of Figure~\ref{fig3}. 
Best-fit parameters are $A=0.099 \pm 0.033$, $\mu = 5.52 \pm 0.18$~MeV and $\sigma = 0.45 \pm 0.14$~MeV and provide a much better agreement to the joint data: $\chi^2/\mathrm{ndf} = 18.8/18$. 
The addition of the best fit bump improves the $\chi^2$ value by 12.0 while reducing the degrees of freedom by 3, corresponding to an excess with significance $2.4 \sigma$  (p-value $0.007$) over the no-bump case. 
Because this comparison incorporates information about the unfolding biases through the $A_C$ matrix, the results do not depend on the method (either Tikhonov or Wiener-SVD) used to perform the joint unfolding.
The deficit of events observed around 7~MeV is driven by a fluctuation in a single bin of \STEREO prompt energy spectrum, as discussed in \cite{STEREOShapePhase2}. 
Due to a strong positive correlation in this energy range, not represented by the diagonal-only error bars in Fig.~\ref{fig3}, the significance of this distortion is small (1.3$\sigma$).

Additionally, a shape-only comparison is made to the deconvolved \uFive{} spectrum from Daya Bay~\cite{PhysRevLett.123.111801} by interpolating the reported spectrum into this analysis' binning and finding a best-fit scaling factor.
Good overall agreement ($\chi^2/\mathrm{ndf} = 21.0/21$) is found between this work and the unfolded \uFive{} spectrum from the Daya Bay collaboration.
This comparison, in combination with the fitted bump size, suggests that \uFive{} contributes to the LEU bump findings, and is consistent with the case of \uFive{} being an equal contributor to the excess.

The analysis presented in this Letter combines the spectral measurements of the two leading HEU experiments, \PROSPECT and \STEREO.
The two measurements, performed with different detector technologies and energy scales, were shown to be in good agreement. 
This joint analysis therefore provides a robust \uFive{} antineutrino energy spectrum. 
The joint spectrum from two separate, validated methods is available for readers to make comparisons using the respective filter matrix, and the Tikhonov result is presented in the Letter.
Comparing to the Huber model shows preference for a bump in the 5-6~MeV region with 2.4$\sigma$ significance.
This result indicates a \uFive{} bump independent of any other isotopes present in LEU reactors.

The supplementary materials include the unfolded joint spectrum, the transformed covariance matrix in antineutrino energy, and the filter matrix $A_C$ encapsulating all unfolding effects, for both the Tikhonov and Wiener-SVD methods.
For quantitative comparisons, readers must apply the filter matrix to any model before comparing to the unfolded data, as it is done in eqn.~(\ref{eqn:gaus-fit}).


The \STEREO experiment is funded by the French National Research Agency (ANR) within the project ANR-13-BS05-0007. Authors are grateful for the technical and administrative support of the ILL for the installation and operation of the \STEREO detector. We further acknowledge the support of the CEA, the CNRS/IN2P3 and the Max Planck Society.

The \PROSPECT experiment is supported by the following sources: US Department of Energy (DOE) Office of Science, Office of High Energy Physics under Award No. DE-SC0016357 and DE-SC0017660 to Yale University, under Award No. DE-SC0017815 to Drexel University, under Award No. DE-SC0008347 to Illinois Institute of Technology, under Award No. DE-SC0016060 to Temple University, under Contract No. DE-SC0012704 to Brookhaven National Laboratory, and under Work Proposal Number  SCW1504 to Lawrence Livermore National Laboratory. This work was performed under the auspices of the U.S. Department of Energy by Lawrence Livermore National Laboratory under Contract DE-AC52-07NA27344 and by Oak Ridge National Laboratory under Contract DE-AC05-00OR22725. Additional funding for the experiment was provided by the Heising-Simons Foundation under Award No. \#2016-117 to Yale University. 

J.G. is supported through the NSF Graduate Research Fellowship Program and A.C. performed work under appointment to the Nuclear Nonproliferation International Safeguards Fellowship Program sponsored by the National Nuclear Security Administration’s Office of International Nuclear Safeguards (NA-241). This work was also supported by the Canada  First  Research  Excellence  Fund  (CFREF), and the Natural Sciences and Engineering Research Council of Canada (NSERC) Discovery  program under grant \#RGPIN-418579, and Province of Ontario.

\PROSPECT further acknowledges support from Yale University, the Illinois Institute of Technology, Temple University, Brookhaven National Laboratory, the Lawrence Livermore National Laboratory LDRD program, the National Institute of Standards and Technology, and Oak Ridge National Laboratory. We gratefully acknowledge the support and hospitality of the High Flux Isotope Reactor and Oak Ridge National Laboratory, managed by UT-Battelle for the U.S. Department of Energy.

\bibliography{./references}{}

\end{document}